\documentclass[10pt,journal,final,twocolumn]{IEEEtran}
\usepackage{dsfont}

\usepackage{subeqnarray}
\usepackage{cleveref}
\usepackage{amssymb}
\usepackage{amsfonts}
\usepackage{amsmath}
\usepackage[dvipdfmx]{graphicx}
\usepackage{bmpsize}
\usepackage{amsmath}
\usepackage[all,poly]{xy}
\usepackage{multirow}
\usepackage{algorithm}
\usepackage{algorithmic}
\usepackage{color}
\usepackage[noadjust]{cite}
\usepackage{ulem}
\usepackage{slashbox}
\title{A Hierarchical Bayesian Approach to Neutron Spectrum Unfolding with Organic Scintillators}
\author{Haonan Zhu, Yoann Altmann, Angela Di Fulvio, Stephen
McLaughlin, Sara Pozzi, Alfred Hero
\thanks{Haonan Zhu and Alfred Hero are with Department of Electrical Engineering and Computer Science, University of Michigan, Ann Arbor MI. Yoann Altmann and Steve
McLaughlin are with School of Engineering and Physical Sciences, Heriot-Watt University, Edinburgh,
U.K.. Angela Di Fulvio is with Department of Nuclear, Plasma, and Radiological Engineering, University of Illinois at Urbana-Champaign, Urbana IL. Sara Pozzi is with Department of Nuclear Engineering and Radiological Sciences
University of Michigan, Ann Arbor, MI.}}

\def\bfm{{\mathbf{m}}}

\def\bfr{{\mathbf{r}}}

\def\bphi{{\mathbf{x}}}
\def\bfy{{\mathbf{y}}}

\def\bfL{{\mathbf{L}}}

\def\bfR{{\mathbf{R}}}

\def\bphi{{\mathbf{X}}}

\newcounter{algo}
\renewcommand{\thealgo}{\arabic{algo}}

\newenvironment{algogo}[1]{
\smallskip
\noindent \hrule\vspace{0.2\baselineskip} \hrule
\begin{small}
\refstepcounter{algo} \center{\bf \textsc{Algorithm \thealgo}}
\\{\center{\bf #1}}
\smallskip
\flushleft
 } {
\end{small}
\smallskip
\hrule\vspace{0.2\baselineskip} \hrule
\smallskip
}

\newcommand{\transp}{^T}
\def\bphi{\boldsymbol{\bf{\phi}}}

\graphicspath{{figures/}}

\begin{document}
\maketitle

%\captionsetup[table]{textfont={sc,footnotesize}, labelfont=footnotesize, labelsep=newline}
\begin{abstract}
 We propose a hierarchical Bayesian model and state-of-art Monte Carlo sampling method to solve the unfolding problem, i.e., to estimate the spectrum of an unknown neutron source from the data detected by an organic scintillator. Inferring neutron spectra is important for several applications, including nonproliferation and nuclear security, as it allows the  discrimination of fission sources in special nuclear material (SNM) from other types of neutron sources based on the differences of the emitted neutron spectra. Organic scintillators interact with neutrons mostly via elastic scattering on hydrogen nuclei and therefore partially retain neutron energy information. Consequently, the neutron spectrum can be derived through deconvolution of the measured light output spectrum and the response functions of the scintillator to monoenergetic neutrons.
The proposed approach is compared to three existing methods using simulated data to enable controlled benchmarks. We consider three sets of detector responses. One set corresponds to a 2.5 MeV monoenergetic neutron source and two sets are associated with (energy-wise) continuous neutron sources ($^{252}$Cf and $^{241}$AmBe). Our results show that the proposed method has similar or better unfolding performance compared to other iterative or Tikhonov regularization-based approaches in terms of accuracy and robustness against limited detection events, while requiring less user supervision. The proposed method also provides a posteriori confidence measures, which offers additional information regarding the uncertainty of the measurements and the extracted information. 
\end{abstract}

\begin{IEEEkeywords}
Organic scintillators; Spectral unfolding; Bayesian Inference; Markov-chain Monte Carlo methods
\end{IEEEkeywords}

\section{Introduction}
\label{sec:introduction}
Two main reactions are exploited in neutron detection: scattering on a light nucleus or capture on elements such as $^6$Li, $^{10}$B or $^3$He. Thermal neutrons (0.025 eV) are preferentially detected via capture reactions because the aforementioned elements exhibit high cross-sections for thermal neutron absorption. Conversely, fast neutrons are detected via scattering reactions on light elements, such as hydrogen and deuterium. The detection of fast neutrons, such as those emitted by SNMs, involves directly exploiting inelastic and elastic scattering reactions, without the need to moderate the source neutrons. Organic scintillators are typically hydrocarbon compounds and detect neutrons via elastic and inelastic scattering reactions on hydrogen nuclei. The energy deposited by scattered proton recoils depends on the scattering angle and it ranges from zero up to the neutron maximum energy. The intensity of light pulses produced by the scintillator is correlated to the energy deposited by the recoil protons \cite{Birks1964}. This light production mechanism allows partial retention of the energy of the impinging neutrons, however, the correlation between the energy of the impinging neutron and the light pulse produced is weak, and therefore deriving the neutron spectrum from the measured data is particularly challenging. 
Finding the energy spectrum of the neutrons impinging on an organic scintillator from its light output response is an ill-posed problem, which often admits multiple solutions \cite{Weise1990305}. This problem is traditionally addressed using so-called unfolding algorithms, which aim at recovering the spectrum that is most likely to have produced the given measured response. 
Accurate unfolding and spectrometry are critical in several applications, such as radiation protection \cite{wiegel2009intercomparison}, nuclear physics \cite{adye2011unfolding}, nonproliferation \cite{lawrence2016warhead} and safeguards \cite{pozzi2010fast}. In safeguards, nonproliferation, and decommissioning applications, accurately discriminating between different neutron sources, such as those based on (alpha, n) reactions and those based on fission, would be a valuable tool when characterizing neutron-emitting samples of unknown composition.

Several parametric unfolding algorithms have been developed over the past decades \cite{root,bedogni2007fruit,matzke1994unfolding,REGINATTO2002242,matzke2002propagation,pehlivanovic2013comparison}. They primarily differ by: 1) the way they model the acquisition process, in particular the distribution of the observation noise, and 2) by the way they combine the knowledge available about the neutron spectrum to be recovered and the measured data. 
Bayesian methods have been previously proposed \cite{weise1989priori,weise1993bayesian,choudalakis2012fully,kuusela2015statistical} in the context of spectrum unfolding. This family of methods aims to regularize ill-posed problems by incorporating prior information about the neutron spectrum to be recovered (denoted as $\bphi$) in a principled way. In this study, we also review other existing approaches  \cite{matzke1985neutron,weise1989priori,weise1993bayesian,puulpan1993unfolding,matzke2002propagation,pehlivanovic2013comparison} and also discuss how they can be (re)interpreted in a Bayesian framework through the use of different prior distributions. With the success of techniques from the artificial intelligence community in a variety of research fields, there has also been an increasing interest in applying such techniques to the unfolding problem. For instance, Artificial Neural Network (ANN) have been applied to recover the neutron spectra \cite{vega2006neutron} when a sufficiently large collection of ground truth spectra is available and can be used as training set of the network. This approach requires a significant amount of prior information (through large sets of reference data) and may fail in analyzing data/samples that are not in line with the training data (e.g., a new source). Heuristic adaptive search-based algorithms, namely genetic algorithms (GA), have also been investigated to obtain the unfolded spectra \cite{freeman1999genetic,suman2014neutron,shahabinejad2016new}, but they do not provide convergence guarantees \cite{sivanandam2008genetic}.  
In this work, we present a Bayesian hierarchical model for neutron unfolding and an associated state-of-art Markov chain Monte Carlo (MCMC) method to infer the unknown neutron spectrum. As it will be shown, the algorithm is able to automatically tune the amount of smoothness  of the recovered spectrum (i.e., how sharply it can vary as the energy changes) at a reduced additional cost. Through several simulation results, we illustrate the potential benefits of our method when compared to traditional approaches.

In order to fairly compare the different algorithms, this paper focuses on simulated data generated using a realistic and widely used Monte Carlo-based simulator of detection events discussed in Section \ref{subsec:mc_simulation}. This approach allows us 
\begin{itemize}
\item to characterize precisely the response function of the detector of interested (EJ-309 here), which would be difficult and extremely time consuming through measurement campaigns; 
\item to obtain simulated detector responses resembling measured ones for known neutron sources (input of the Monte Carlo simulator); 
\item to avoid signal distortion caused by potential experimental limitations (e.g., imperfect material shielding or room returns).
\end{itemize}
In this work, we simulated the response of the detector to three types of sources: a 2.5 MeV monoenergetic one, which can be obtained from the measurement of a deuteron-deuteron fusion reaction, an $^{241}$AmBe ($\alpha$, n) spectrum and a $^{252}$Cf fission spectrum.  

The remainder of the paper is organized as follows. Section \ref{sec:method} introduces how the simulated data have been generated using a semi-empirical model and Monte Carlo simulation. This section also reviews briefly the main existing unfolding methods as well as the proposed method. The obtained results and a quantitative comparison between the unfolding methods are presented and discussed in Section \ref{sec:results}. Conclusions are finally reported in Section \ref{sec:conclusion}.

\section{Methods}
\label{sec:method}

\subsection {Organic Scintillator Response and Monte Carlo Simulation}
\label{subsec:mc_simulation}

Scintillators emit light upon interaction with ionizing radiation. Organic scintillators are compounds of hydrogen and carbon, and are suitable to detect fast neutrons. Neutron elastic scatter on a hydrogen nucleus produces a scattered neutron and a recoil proton. In the energy range of interest ($<20$ MeV neutrons), it can be assumed that the recoil proton deposits all its energy within a detector of practical size, e.g. 7.62-cm diam. by 7.62-cm length. 
%Fluorescence is the emission of light pulses following the molecular excitation caused by radiation interaction and ultimately allows its detection \cite{Birks1964}. 
%\sout{Light readout is typically performed using a photomultiplier vacuum tube (PMT), which generates a current pulse in response to a light signal from the scintillator. The voltage drop resulting from the current transient is the detected pulse. The height and the integral of the pulses are proportional to the energy deposited in the crystal by one or multiple interactions.}
The light output response is approximately linear with the energy deposited by electrons, $E_e$, with energy above approximately 40 keV \cite{williamson1999plastic}. Therefore, the detector light output is conveniently expressed in terms of electron
light output ($ee$: electron-equivalent  units). In practice, the upper edge of the known Compton electron  distribution produced by a monoenergetic gamma-ray source, e.g. $^{137}$Cs, provides a suitable calibration point, commonly referred to as the Compton edge, $V_{CE}$. The light output in electron equivalent units ($y_{ee}$) is therefore calculated at any pulse height voltage $V$ as in Eq. \eqref{eq:2.3}. 
\begin{equation}
\label{eq:2.3} 
y_{ee} =  \frac{\bar{E}_{ee}}{V_{CE}} V.
\end{equation}
In equation \eqref{eq:2.3}, $\bar{E}_{ee}$ is the maximum energy deposited by a Compton-recoil electron, in electron-equivalent energy units. Conversely, the light output response to charged particles heavier than electrons, like neutron-produced recoil protons, is not linear with the energy deposited. Throughout this paper, $y$ identifies the light output in electron-equivalent energy units, e.g., $keV_{ee}$. A widely accepted set of models which semi-empirically describes the dependence of the light output $y$ with the proton energy deposited $E_p$ and the energy deposited-per-unit-length $dE_p/dx$ was first introduced by Birks \cite{Birks1964} and is reported in Eq. \eqref{eq:2.4} below
\begin{equation} 
\label{eq:2.4} 
y(E'_p) = \int_{0}^{E'_p} \frac{S \: dE_p}{(1+k_B\: dE_p/dx)} .
\end{equation}
Equation \eqref{eq:2.4} is the integral over energy of Eq. (3) in the paper by Brooks \emph{et al.} \cite{Brooks1979477}. In Eq. \eqref{eq:2.4}, $S$ is the scintillation efficiency, in $MeV_{ee}$, and $k_B$ is a material-dependent constant, in $g/MeV cm^{2}$ , often referred to as the Birks' coefficient \cite{Brooks1979477}. We simulated the pulse height distributions, i.e. light output spectra, of a 7.62-cm diam by 7.62 length EJ-309 detector in response to monoenergetic neutrons, for 500 evenly distributed neutron sources with energy between 0.1 MeV to
20 MeV, using MCNPX-PoliMi \cite{pozzi2012mcnpx}. 
We used MPPost, a MCNPX-PoliMi post-processing code, to obtain the light output spectrum %\sout{convolve the detector light output response with the simulated neutron fluence and produce the corresponding light output spectrum}
, i.e. the frequency of occurrence of pulse amplitudes in a given measurement time \cite{Miller}.
An enhanced version of MPPost allows the use of the semi-empirical model in equation \eqref{eq:2.4} to generate the detector-specific light output spectrum \cite{Norsworthy201720}. For EJ-309, the coefficients $S$ and $k_B$ that we used are $2.277$ MeVee/MeV and $33.84$ g/MeV cm$^2$, respectively \cite{Norsworthy201720}. The software also applies a Gaussian smear to account for the detector's energy resolution. The energy resolution function that we implemented was measured by Enqvist et al. \cite{ENQVIST201379} for the type of detector under investigation and is reported in Eq. \eqref{eq:2.5}, where $a=0.113\pm0.007$, $b=0.065\pm0.011 MeV^2$, and $c=0.060\pm0.005 MeV$.
\begin{equation} 
\label{eq:2.5} 
(\Delta E/E) = (\sqrt{a^2 + b^2/E + (c/E)^2})
\end{equation}

\begin{figure}[h!]
\centering
\includegraphics[width=\columnwidth]{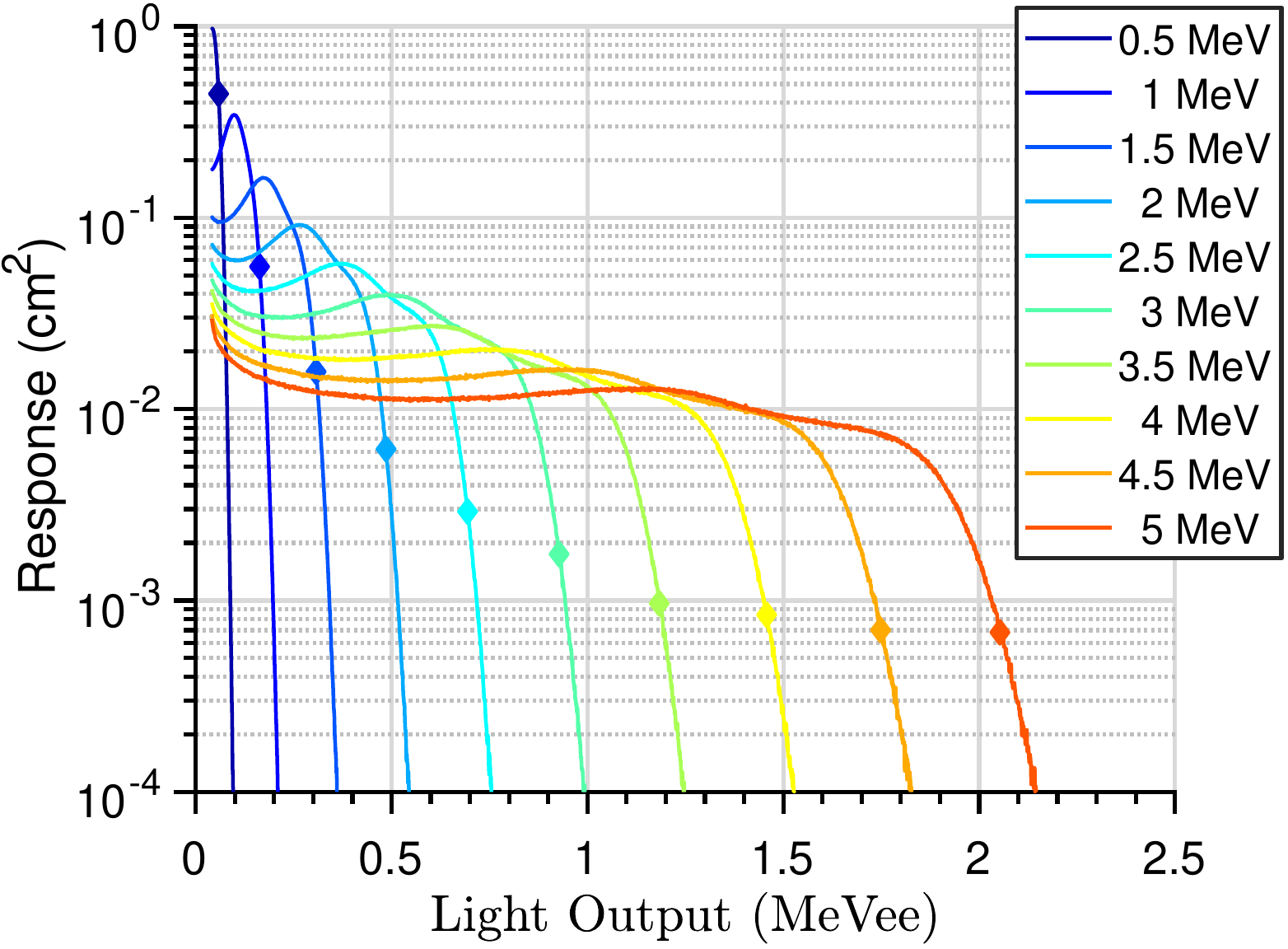}
\caption{Simulated Response Functions for a 7.26 cm diam. by 7.26 cm length EJ-309 detector in response to monoenergetic neutrons in the 0.5-5 MeV range. The solid diamonds show the light output corresponding to the maximum energy deposited.}
\label{fig:1}
\end{figure}

Fig. \ref{fig:1} shows the simulated light output spectra produced by irradiation with selected mono-energetic neutron sources between 0.5 MeV and 5 MeV. %\sout{ One may notice that the response of the detector to a monoenergetic neutron does not correspond to a single line in the light output spectrum. The response to monoenergetic neutrons is instead continuous up to the maximum energy deposited and is determined by the physics of the neutron scattering reaction.}

The energy deposited in the detector by recoil protons $E_p$ after elastic collision with neutrons of energy $E$ depends on the scattering angle of the charged recoil in the laboratory system of reference: $\theta$ (see Eq. \eqref{eq:2.6}).

\begin{equation} \label{eq:2.6} 
E_p = \frac{4A}{(1+A)^2} \: cos^2  \theta \: E 
\end{equation}

In the elastic scattering kinematics equation (Eq. \eqref{eq:2.6}), A is the mass number of the target nucleus (A=1 for $^1$H). Monoenergetic neutrons can thus produce proton recoils in the energy range from $E_{pmax} = E$, when $\theta = 0$, to zero, when $\theta = \frac{\pi}{2}$ and consequently light pulses with amplitude ranging form $y(E_{pmax})$ to $0$. Note that in Fig. \ref{fig:1}, the light  output corresponding to the maximum energy deposited by proton recoils is identified by solid diamonds. We determined this light-output value as the minimum of the derivative of the upper edge of the light output spectrum, following the same method proposed by Kornilov and colleagues \cite{Kornilov2009226}.

\begin{figure}[h!]
\centering
\includegraphics[width=\columnwidth]{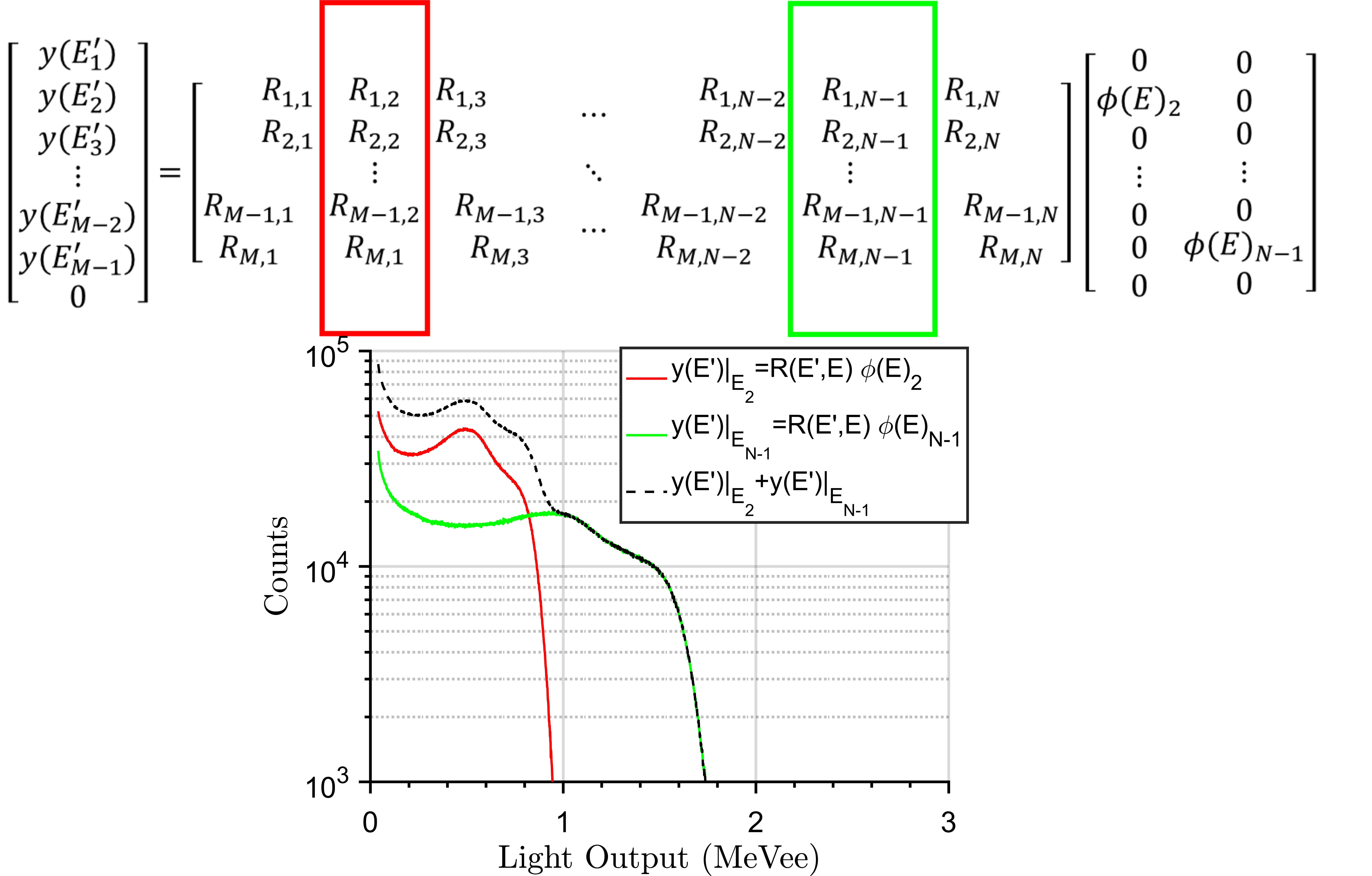}
\caption{Example of the convolution between an ideal neutron spectrum with two energy peaks and the detector response matrix.}
\label{fig:2}
\end{figure}

As in any spectroscopy-capable sensor, the number of counts at a given bin of the light output spectrum $y (E')$ ($E'$ in $ee$) is given by the convolution of the detector response at that light output bin with the impinging neutron spectrum, as formalized in the next section (Eq. \eqref{eq:2.1}). 
Fig. \ref{fig:2} shows the process of spectrum unfolding for two monoenergetic neutron spectra on discretized data sets. One may notice that an ideal monoenergetic neutron spectrum is a linear transformation of one element of the canonical basis for the response matrix and therefore selects only one corresponding light output response, i.e. column of the response matrix. For organic scintillation detectors, the number of neutron energy bins ($N$) is of the same order of magnitude as the number of light-output channels measured ($M$). In neutron spectroscopy, this case is usually referred to as multi-channel unfolding, as opposed to few-channel unfolding, where $M << N$. Few-channel unfolding applies to other types of detectors, e.g. Bonner spheres \cite{BEDOGNI2002381} and superheated emulsions \cite{d2009angle}.
The size of the response matrix used in this work is $600 \times 149$ (i.e., $M=600$ and $N=149$). These channel numbers correspond to a light output bin width of 0.001 MeVee, in the 0.01-6 MeVee light-output range, and a neutron energy bin width of 100 keV, in the 0.1-15 MeV energy range.

\subsection {Discretized observation model}
The detector response function is denoted by $R(E',E)$. More precisely, $R(E',E_0)$ is the light output spectrum (with $E'$ in eVee) in response to a monoenergetic neutron of energy $E_0$.  The light output and unknown neutron energy spectral fluence, i.e. the number of neutrons per unit area \cite{ISO8529}, also referred to as neutron spectra throughout this paper, are related through the following Fredholm integral equation \cite{weise1989priori,puulpan1993unfolding,weise1993bayesian,matzke1985neutron,matzke2002propagation}
\begin{equation} \label{eq:2.1}
y(E')=\int_0^{\infty}R(E',E)\phi(E)dE.
\end{equation}
For numerical computation, Eq. \eqref{eq:2.1} can be approximated by the following linear equation
\begin{equation} \label{eq:2.2}
\bfy  \approx \bfR \bphi,
\end{equation}
where $\bphi = [\phi_1,\ldots,\phi_N]\transp \in \mathbb{R}_{+}^N$ denotes the neutron spectrum discretized over $N$ energy bins, $\bfy=[y_1,\ldots,y_M]\transp \in \mathbb{R}_{+}^M$ is light output spectrum discretized over $M$ bins and $\bfR$ is the $M \times N$ response matrix of the detector. 
Unfolding methods aim at recovering $\bphi$ from $\bfy$ such that Eq. \eqref{eq:2.2} is satisfied. However, they can differ by the similarity measures or likelihood functions used to compare $\bfy$ and $\bfR \bphi$.
A classical approach to matching $\bfy$ and $\bfR \bphi$ consists of considering a quadratic similarity measure
\begin{eqnarray}
\label{eq:LS}
||\bfy - \bfR\bphi||_{\boldsymbol{\Sigma}}^2= (\bfy - \bfR \bphi)\transp\boldsymbol{\Sigma}^{-1}(\bfy - \bfR \bphi),
\end{eqnarray}
where the $M \times M$ matrix $\boldsymbol{\Sigma}$ relates to the characteristic of the measurement noise. If $\boldsymbol{\Sigma}$ is set to the identity matrix, Eq. \eqref{eq:LS} reduces to the classical least-squares criterion $||\bfy - \bfR\bphi||_{2}^2$ where $||\cdot||_2$ denotes the standard $\ell_2$ norm. Recovering $\bphi$ using the criterion in Eq. \eqref{eq:LS} implicitly assumes that $\bfy$ is a noisy version of $\bfR \bphi$ corrupted by Gaussian noise with covariance matrix (proportional to) $\boldsymbol{\Sigma}$, i.e., 
\begin{eqnarray}
\label{eq:lik_Gaussian}
\bfy | \bphi \sim \mathcal{N} \left(\bfR \bphi, \boldsymbol{\Sigma} \right),
\end{eqnarray}
where $\bfy | \bphi$ reads "$\bfy$ given $\bphi$",  $\sim$ reads "is distributed according to" and $\mathcal{N} \left( \bfm, \boldsymbol{\Sigma} \right)$ denotes the multivariate Gaussian distribution with mean $\bfm$ and covariance matrix $\boldsymbol{\Sigma}$. Indeed, it can be easily shown that minimizing \eqref{eq:LS} with respect to (w.r.t.) $\bphi$ is equivalent to maximizing the likelihood \eqref{eq:lik_Gaussian} w.r.t. $\bphi$, as will be discussed in the next section. 

Since the acquisition process consists of detecting individual neutrons (discrete number of events within a given time period), it is reasonable to consider Poisson noise models. These models enable the consideration of the correlation between the mean (expected) detection rates and the variance of the observation noise. Moreover, such models are more suited for low counts (e.g. less than 10 per bin), as investigated in Section \ref{sec:results} where we consider scenarios with as few as 1 count per light output bin on average. The classical Poisson noise model assumes that the light output in the $M$ energy bins are mutually independent and Poisson distributed. The resulting observation model becomes \cite{choudalakis2012fully}
\begin{eqnarray}
\bfy | \bphi \sim \mathcal{P} \left(\bfR \bphi \right),
\end{eqnarray}
where $\mathcal{P} \left( \cdot \right)$ denotes the element-wise Poisson distribution, i.e., $\forall m, y_m|\bphi \sim \mathcal{P} \left( \bfr_{m,:}\bphi \right)$ with $\bfr_{m,:}$ the $m$th row of $\bfR$. Consequently, the likelihood of the observed light output spectrum $\bfy$ given the underlying neutron spectrum $\bphi$, denoted $f(\bfy|\bphi)$ can be expressed as 
\begin{eqnarray}
f(\bfy|\bphi) = \prod_{m=1}^M \dfrac{(\bfr_{m,:}\bphi )^{y_m}}{y_m!} \exp \left[ -\bfr_{m,:}\bphi \right].
\end{eqnarray}

In this subsection, we have discussed how the unfolding problem can be
formulated as a linear inverse problem and discussed two main noise observation models. In the next subsection, we review the primary existing unfolding methods and their relation with the observation models discussed above. These methods will then be used in Section \ref{sec:results} to assess the performance of the proposed approach.

\subsection {Existing unfolding approaches}
\label{subsec: existing_app}
% \subsubsection{Maximum-likelihood based algorithms}
% \label{subsubsec:mle}
The first statistical approach to unfolding is a classical method for inverse problems and is referred to as Maximum Likelihood Estimation (MLE). MLE-based unfolding recovers the neutron spectrum by finding $\bphi$ that maximizes the likelihood function \cite{CaseBerg:01}. Maximizing the likelihood $f(\bfy|\bphi)$ is equivalent to minimizing the negative log-likelihood, (which is often preferred for algorithmic stability since $-\log\left(f(\bfy|\bphi)\right)$ is often a (nearly) quadratic function). Although we can consider as many MLE-based algorithms as likelihood models, we primarily focus on Gaussian and Poisson noise models here. More precisely, using an isotropic Gaussian noise model is equivalent to  using a classical minimization of least square loss, while the Poisson model is preferred for counting data as discussed above. Under Poisson noise assumption, the log-likelihood reduces to%preliminary results have led to generally better results using MLE based on Poisson noise rather than isotropic Gaussian noise. Thus, here we only consider the former estimation strategy.\\

$\log(f(\bfy|\bphi))$
\begin{eqnarray}
\label{eq:log_lik_Poisson}
 = \sum_{m=1}^M y_m\log(\bfr_{m,:}\bphi )-\log{(y_m!)} -\left(\bfr_{m,:}\bphi \right).
\end{eqnarray}
%Several approaches can be used to maximize \eqref{eq:log_lik_Poisson}. Among them, the Expectation - Maximization (EM) algorithm will approximate the solution iteratively using the following iterative rule \cite{dempster1977maximum,pehlivanovic2013comparison}
%\begin{equation}  
%\phi^{k+1}_{j} = \frac{\phi^k_j}{\sum_{m=1}^{M} r_{m,j}}\sum_{m=1}^M\frac{y_m}{\bfr_{m,:}\bphi^{k}}, \forall j \in 1, \ldots N.  
%\end{equation}%

Maximum likelihood estimation aims at recovering the unknown spectrum from the data only, i.e., without additional information), by inverting (or pseudo inverting) the response matrix and using a cost function accounting for the statistical properties on the observation noise. This is a simple inference strategy but can provide poor results in the presence of noise, especially when the response matrix is ill-conditioned (as it is often the case in practice). Thus, maximum penalized likelihood estimation methods based on Poisson likelihood models have been proposed. Since we expect most of the unknown neutron spectra to be recovered are relatively smooth, it makes sense to add a regularization which reflects this prior belief. Here we chose a regularization term that promotes small second-order derivative (in the spectral dimension), which results in the following objective function to be minimized
\begin{eqnarray}
\label{eq:penalized_mle}
\underset{\bphi \in \mathcal{R}_+^N}{\min} \sum_{m=1}^M - \log(f(\bfy|\bphi)) + \lambda ||\bfL\phi||_2^2,
\end{eqnarray}
\noindent where $\lambda$ is a tuning parameter that controls the smoothness, $\log(f(\bfy|\bphi))$ is defined in \eqref{eq:log_lik_Poisson} and $\bfL$ denote the discrete Laplace operator, which can be written as
\begin{equation}
\label{eq:lap_matrix}
\bfL =
\begin{pmatrix}
-2 & 1 & 0 & \cdots & \cdots & \cdots & \cdots & 0\\
1 & -2 & 1 & 0 & & & & \vdots\\
0 & 1 & -2 & 1 & \ddots & & & \vdots\\
\vdots & 0 & \ddots & \ddots & \ddots & \ddots & & \vdots\\
\vdots & & \ddots & \ddots & \ddots & \ddots & 0 & \vdots\\
\vdots & & & \ddots & 1 & -2 & 1 & 0\\
\vdots & & & & 0 & 1 & -2 & 1\\
0 & \cdots & \cdots  & \cdots & \cdots & 0 & 1 & -2\\
\end{pmatrix}.
\end{equation}

There are multiple ways of solving the minimization problem in Eq. \eqref{eq:penalized_mle}, e.g., using Alternating Direction Method of Multipliers (ADMM) \cite{boyd2011distributed} as in Poisson image deconvolution by augmented Lagrangian (PIDAL) (see \cite{figueiredo2010restoration}) or using sequential Gaussian approximations of the Poisson likelihood \cite{harmany2010spiral}. Here, we chose the ADMM implementation presented in \cite{figueiredo2010restoration} for its simplicity and relatively low computational cost. It is worth noting that the One-Step-Late (OSL) algorithm in \cite{10.2307/2345668,pehlivanovic2013comparison} is an alternative method to approximate the solution of Eq. \eqref{eq:penalized_mle}. Note that Eq. \eqref{eq:penalized_mle} requires to select an appropriate value of $\lambda$, which will affect the quality of the solution. This point will be further discussed in Section \ref{sec:results}.

Under the Gaussian noise model, the unfolded spectrum is a solution to the convex optimization problem as in \eqref{eq:penalized_mle} where $- \log(f(\bfy|\bphi))$ is replaced with the standard quadratic loss function $||\bfy - \bfR\bphi||_{2}^2$.  The non-negativity constraints imposed on the unfolded spectrum prevent us from having a closed form solution, thus we applied an ADMM algorithm with L-curve method \cite{hansen1999curve} to obtain the unfolded spectrum. This algorithm will be referred to Tik (Tikhonov Regularizer) in remainder of the paper.

Among the methods whose codes are available, we also used GRAVEL presented in \cite{matzke1994unfolding,MCunfoldingProgram}. The iterative update rule of GRAVEL algorithm (at iteration $(k+1)$) is given by
% \begin{equation*}
% \label{GRAVEL}
% \bphi_{n}^{(k+1)}=\bphi_{n}^{(k)} \exp \left(\frac{\sum_{j} W_{n j}^{(k)} \log\left(\frac{\bfy_{j}}{\sum_{i} R_{j i} \bphi_{i}^{(k)}}\right)}{\sum_{j} W_{n j}^{(k)}}\right)    
% \end{equation*}
\begin{equation}
\label{GRAVEL}
\phi_{n}^{(k+1)}=\phi_{n}^{(k)} \exp \left(\frac{\sum_{m} W_{n,m}^{(k)} \log\left(\frac{y_{m}}{\bfr_{m,:} \bphi^{(k)}}\right)}{\sum_{m} W_{n
m}^{(k)}}\right), \forall n,
\end{equation}
where $\bphi^{(k)}$ is estimated neutron spectrum at iteration $k$, $\sigma_m$ is an estimate of measurement error in the $m$th light output bin, $r_{m,n}=[\bfR]_{m,n}$ and 
\begin{eqnarray}
W_{n,m}^{(k)}=\frac{r_{m,n} \bphi_{n}^{(k)}}{\sum_{i} \left(r_{m,i} \bphi_{i}^{(k)}\right)} \frac{y_{m}^{2}}{\sigma_{m}^{2}}
\end{eqnarray}

GRAVEL allows the user to incorporate prior information, when available, as an a priori known default spectrum. We have used a flat spectrum for consistency with the other methods. Regardless of the type of source, a flat initial spectrum was used, whose boundaries are detailed in Table \ref{tab:1}. The spectrum intensity had a negligible impact on the final results. The boundaries of the light output spectra are reported in Table \ref{tab:1} and vary according to the simulated data. Light-output bins with a relative statistical error higher than $20\%$ in the high-energy tail of the light output spectra were excluded. The uncertainty associated with the simulated bins was calculated as the square root of the counts. 
%The maximum number of iterations was increased until the relative fluctuation in the chi-squared per degree of freedom (PDF) was below $0.0004\%$.
GRAVEL stopping criterion is either the user-defined chi-squared per degree of freedom (PDF) or the input maximum number of iterations (to stop the algorithm after a given number of iterations if the first criterion is not satisfied yet) \cite{matzke1997unfolding}. 
%The chi-squared parameter is the sum of the squares of the deviations between the simulated light output spectrum and the light-output spectrum obtained by folding the unfolded spectrum with the response matrix, divided by the standard uncertainties of the measured light output. 
In our case, the number of degrees of freedom is $M$ and the chi-squared-PDF was set to one, while the maximum number of iterations was 6000. For the $^{252}$Cf and $^{241}$AmBe spectra (see Section \ref{sec:results}), the algorithm reached the desired chi-squared PDF after few iterations ($<20$), while the maximum number of iterations criterion was adopted for the monoenergetic spectrum, for which the relative fluctuation in the chi-squared PDF was below $0.0004\%$, after 6000 iterations.
The GRAVEL parameters used in Section \ref{sec:results} are reported in Table \ref{tab:1}.

\begin{table}[h!]
%\begin{center}
\resizebox{1\columnwidth}{!}{%
\begin{tabular}{|c|c|c|c|}
\hline
 Parameters & $^{241}$AmBe & $^{252}$Cf & 2.5 MeV  \\
\hline
$LO_{min}$-$LO_{max}$ (MeVee) & 0.05-5.8 & 0.05-4.2 & 0.05-0.83\\
\hline
$E_{min}$-$E_{max}$ (MeV) & 0.5-15.0 & 0.5-15.0 & 0.5-3.0\\
\hline
%Set Chi-squared PDF & 1 & 1& 1\\
%\hline
%Final Chi-squared PDF & 0.98 &  0.99 & 24 \\
%\hline
\end{tabular}
}
\vspace{0.3cm}
\caption{Specific parameters and settings used to unfold the neutron spectra in GRAVEL.}
%\end{center}
\label{tab:1}
\end{table}

MAXED is another unfolding computer program available within the UMG package \cite{REGINATTO2002242}.  MAXED applies the maximum entropy principle to the deconvolution of spectrometer data. The obtained results were similar to those calculated using GRAVEL, therefore MAXED was not included as an additional comparison methods. 
\subsection{Novel Bayesian spectrum unfolding approach}
\label{subsec: proposed_approach}

Bayesian methods have been previously proposed \cite{weise1989priori,weise1993bayesian,choudalakis2012fully,kuusela2015statistical,reginatto2008bayesian} in the context of spectrum unfolding. As mentioned earlier, they aim at regularizing ill-posed problems by incorporating a-priori information about $\bphi$ in a principled way. More precisely, such knowledge is incorporated through a so-called prior distribution $f(\bphi|\delta)$, parameterized by $\delta$. The selection of the prior distribution $f(\bphi|\delta)$ is guided by the amount of prior information available and the induced algorithm complexity \cite{choudalakis2012fully}. Moreover, the choice of this distribution can be crucial when the amount of information contained in the data in limited, e.g., in the presence of few observations and noisy data. While informative prior distributions will greatly improve the estimation performance if appropriately tailored, they will negatively impact the estimation performance if the data deviates from the the prior belief. In previous studies \cite{weise1989priori,weise1993bayesian}, empirical Bayes methods were used, in which the prior distribution was built from previously acquired data. However, such methods perform poorly if the neutron spectrum to be recovered is not in agreement with the data-driven prior distribution. Bayes' theorem provides a formal way to combine our prior belief $f(\bphi|\delta)$ with the observations (through the likelihood $f(\bf{y}|\bphi)$) to obtain and exploit $f(\bphi|\bfy,\delta)$. This so-called posterior distribution is classically exploited using summary statistics, including various Bayesian point estimators such as the widely used maximum a posterior (MAP) estimator \cite{weise1989priori,weise1993bayesian} (which can also be seen as maximum penalized likelihood estimation) and posterior means (as in \cite{choudalakis2012fully}) and a posteriori measures of uncertainty (e.g., confidence regions). However, the posterior distribution (e.g. its mode or mean) can highly depend on the value of $\delta$. A classical approach thus consists of incorporating this parameter in the estimation process by extending the Bayesian model and designing an additional prior distribution $f(\theta)$. Applying the Bayes' rule to that model leads to 
\begin{equation} 
\label{eq:Bayes}
f(\bphi,\delta|\bfy)=\frac{f(\bfy|\bphi)f(\bphi|\delta)f(\delta)}{f(\bfy)}\propto f(\bfy|\bphi)f(\bphi|\delta)f(\delta), 
\end{equation}
where the posterior distribution $f(\bphi,\delta|y)$ summarizes the complete information available about $(\bf{\phi},\delta)$, having observed $\bfy$. 

In a similar fashion to the penalized likelihood method in \eqref{eq:penalized_mle}, we choose to assume that the unknown neutron spectrum to be recovered presents smooth variations across neighboring energy bins. This is achieved by assigning $\bphi$ a truncated multivariate Gaussian distribution 
\begin{eqnarray}
\bphi|\delta \sim \mathcal{N}_{\mathbb{R}^+}(\bf{0},\delta\Sigma),
\end{eqnarray}
to ensure the non-negativity of $\bphi$. In this work, we chose $\Sigma^{-1}=\bfL\transp \bfL $, where $\bfL$ is defined as in \eqref{eq:lap_matrix} and the overall amount of smoothness of the solution is governed by the parameter $\delta$ (in a similar fashion to $\lambda$ in the ADMM algorithm). The smaller $\delta$, the smoother the solution. Note that if $\delta$ is fixed (which is not the case here), the solution of PIDAL is obtained using MAP estimation.% But this precision matrix is singular, which leads to an improper prior distribution. 

As shown in Eq. \eqref{eq:Bayes}, we do not choose a fixed value of $\delta$ but assigned to it an inverse-gamma conjugate prior distribution, i.e., $\delta \sim \mathcal{IG}(\alpha_1,\alpha_2)$ with $(\alpha_1,\alpha_2)$ fixed and selected based on WAIC (Watanabe-Akaike Information Criteria) \cite{gelman2014understanding}. %Notice here the proposed prior is weakly informative but leads to a proper conditional distributions as shown in \ref{eq:post_theta}. 
Since in practice $N$ is large, $f(\phi|\delta)$ dominates $f(\delta)$ (as noted in Chapter 4 of \cite{gelman2014bayesian}) and the prior distribution $f(\delta)$ has a limited impact on the estimated neutron spectrum. Moreover, as will be shown in the next paragraph, the conjugacy between $f(\bphi|\delta)$ and $f(\delta)$ will also simplify the estimation procedure.

To exploit the posterior distribution $f(\bphi,\delta|\bfy)$, in this work we apply a Markov chain Monte Carlo (MCMC) method which consists of generating random variables distributed according to $f(\bphi,\delta|\bfy)$. The generated samples are then used to approximate the posterior mean of $\bphi$ and associated a posteriori uncertainty intervals. The pseudo-code of the proposed method is summarized in Algo. \ref{algo:HMC}. 

The proposed approach is similar to the work in \cite{kuusela2015statistical} in the sense that we are also using MCMC methods to solve the unfolding problem. However, several important differences can be highlighted. First, as in \cite{kuusela2015statistical}, we estimate the regularization parameters $\delta$, but this is achieved here through a hierarchical Bayesian model (prior distribution assigned to $\delta$) which yields a more computationally efficient algorithm (fewer iterations required) while this parameter is estimated via maximum marginal likelihood estimation in \cite{kuusela2015statistical}. This approach allows us to also account for the fact that $\delta$ is unknown and the additional uncertainty is automatically included when computing confidence regions for $\bphi$. Second, here we use a constrained Hamiltonian Monte Carlo methods (as discussed below) which improves the sampler convergence and mixing properties compared to traditional sequential Gibbs updates and random walk-based Metropolis-Hastings updates (as in \cite{kuusela2015statistical}).

\begin{algogo}{HMC unfolding algorithm}
\label{algo:HMC}
 \begin{algorithmic}[!h]
			\STATE \underline{Fixed input parameters:} $(\alpha_1,\alpha_2),\sigma^2$, number of burn-in iterations $N_{\textrm{bi}}$, total number of iterations $N_{\textrm{iter}}$.
			\STATE \underline{Initialization ($k=0$)}
			\STATE Set $\bphi^{(0)}={\bf 1},\delta^{(0)}=\alpha_2/(1+\alpha_1)$
			\FOR{$k=1,\ldots N_{\textrm{iter}}$}
			\STATE Sample $\bphi^{(k)} \sim f(\bphi|\bfy,\delta^{(k)})$ using HMC
			\STATE Sample $\delta^{(k)} \sim f(\delta|\bfy,\bphi^{(k)})$ from \eqref{eq:post_theta}
\ENDFOR
		\STATE Set $\widehat{\bphi}=1/(N_{\textrm{iter}}-N_{\textrm{bi}})\sum_{k=N_{\textrm{bi}}+1}^{N_{\textrm{iter}}} \bphi^{(k)}$
\end{algorithmic}
\end{algogo} 

Sampling from $f(\bphi,\delta|\bfy)$ is achieved by sampling iteratively from $f(\bphi|\bfy,\delta)$ and $f(\delta|\bfy,\bphi)$ (lines 5 and 6 of Algo. \ref{algo:HMC}). It can be easily shown using $f(\delta|\bfy,\bphi)\propto f(\bphi|\delta)f(\delta)$ that 
\begin{eqnarray}
\label{eq:post_theta}
\delta|(\bfy,\bphi) \sim \mathcal{IG}\left(\frac{N}{2}+\alpha_1,\frac{\bphi\transp\Sigma^{-1}\bphi}{2} +\alpha_2\right), 
\end{eqnarray}
which is straightforward to sample from. The conditional distribution $f(\bphi|\bfy,\delta)$ is a non-standard distribution and accept/reject
procedures are required to update $\bphi$. Due to the potentially large dimensionality of $\bphi$ (large number $N$ of bins) and the high correlation between these variables, we resort to a constrained Hamiltonian Monte Carlo (HMC) update which uses the local curvature of the distribution
$f(\bphi|\bfy,\delta)$ to propose candidates in regions of high probability. This approach allows better mixing properties than more
standard random walk alternative strategies. The interested reader is invited to consult \cite{Brooks2011} for additional details about Hamiltonian Monte Carlo sampling and \cite{Altmann2017} for an example of application to linear inverse problems involving Poisson noise. The marginal posterior
mean $\widehat{\bphi}$ is approximated by averaging the generated variables after having removed the first $N_{\textrm{bi}}$ iterations of the
sampler which correspond to the burn-in period of the sampler. Similarly, the marginal 95\% credible interval for each $\phi_n$ is computed from the generated samples $\{\phi_n^{(k)}\}_{k}$. The duration of the transient period $N_{\textrm{bi}}$ and the total number of
iterations $N_{\textrm{iter}}$ are set by visual inspection of the chains from preliminary runs. These values are then kept unchanged throughout all the experiments. Note that as mentioned above, by embedding $\delta$ in the Bayesian model through $f(\delta)$ and sampling from $f(\bphi,\delta|\bfy)$, the posterior mean and confidence regions already account for the fact that $\delta$ is unknown (they are computed according to $f(\bphi|\bfy)$). For completeness, the main parameters of the TiK, PIDAL, and MCMC algorithms are summarized in Table \ref{table_2} below, while the settings used for the three different sources in GRAVEL have been already introduced in Table \ref{tab:1}.

\begin{table}[h!]
%\begin{center}
\resizebox{1\columnwidth}{!}{%
\begin{tabular}{|c|c|c|c|}
\hline
Method & Nb. of parameters & Parameters & Value(s)  \\
\hline
Tik & 1& $\lambda$& L-curve \cite{hansen1999curve} \\
\hline
%GRAVEL & 4 & 
%        \begin{tabular}{@{}c@{}c@{}c@{}}$E_{min}$-$E_{max}$ \\ $LO_{min}$-$LO_{max}$ \\ n. iterations \\ default spectrum \end{tabular} 
%        & \begin{tabular}{@{}c@{}c@{}c@{}}0.5-15 MeV \\ 0.1-6.0 MeVee \\ 6,000 \\ flat spectrum \end{tabular} \\
% \hline
PIDAL & 1 & $\lambda$ & user-defined\\
\hline
MCMC & 2 & $(\alpha_1,\alpha_2)$& using \cite{gelman2014understanding}\\
\hline
\end{tabular}
}
\vspace{0.3cm}

\caption{Parameters and settings used to unfold the neutron spectra.}
%\end{center}
\label{table_2}
\end{table}

\section{Unfolding Results and Discussion}
\label{sec:results}

We assess the performance of proposed algorithm (referred to as MCMC in the remainder of the paper) with GRAVEL \cite{matzke1994unfolding,matzke1997unfolding,chen2014unfolding},  Tik (Tikhonov regularization with L-curve method) \cite{hansen1999curve} and PIDAL \cite{figueiredo2010restoration} applied to simulated neutron sources. We consider three sources: 2.5 MeV monoenergetic neutron source, $^{252}$Cf and $^{241}$AmBe. The data simulation has been performed using the Monte Carlo method detailed in Section \ref{subsec:mc_simulation} that takes into account the physical process of light output detection with a total number of $5.10^7$ detection events, and we use the semi-empirical response matrix described in Section \ref{subsec:mc_simulation} to unfold the measured light output. In the following experiments, we use the precision matrix $\Sigma^{-1}=\bfL\transp \bfL $ as discussed in Section \ref{subsec: proposed_approach} for the MCMC algorithm and Tik to be consistent with the PIDAL algorithm. In this paper, we select the optimal (in the sense of the performance measure in Eq. \eqref{eq:SAM}) smoothing parameter of PIDAL based on the ground truth, and the resulting method is denoted as PIDAL-O, which stands for oracle PIDAL, in the sense that this approach uses the value of the smoothing parameter which gives the best reconstruction performance, which is in practice impossible to obtain without knowing the spectrum to be recovered. This method assumes access to the ground truth spectra, so it can be seen as the optimal MAP estimator and serves as a way to evaluate the difficulty of the unfolding problem.    

\begin{figure}[h!]
\centering
\includegraphics[width=1\columnwidth]{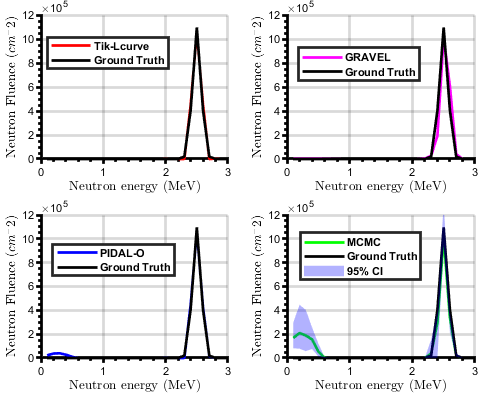}
\caption{Examples of unfolded spectra of the simulated 2.5 MeV monoenergetic neutron source ($5.10^7$ detection events per light output spectrum). MCMC provides additional uncertainty evaluation through credible intervals (CIs), defined here as the high density regions that contain $95\%$ of the samples drawn from the full posterior distribution (leaving $2.5\%$ on each side). Note PIDAL-O (PIDAL-Oracle) assumes full knowledge about ground truth spectra, so it serves as an estimate of the optimal unfolding algorithm and it is not attainable in actual experimental settings.}
\label{fig:3}
\end{figure} 

Fig. \ref{fig:3} shows the unfolded spectra obtained by Tik, GRAVEL, PIDAL-O and MCMC for the simulated 2.5 MeV monoenergetic neutron source. All methods are able to identify the intensity of the peak. MCMC provides additional uncertainty quantification tools through a posteriori Credible Interval (CI). Here we used a $95\%$ CI corresponding to the high density region that contains $95\%$ of the samples drawn from the full posterior distribution (leaving $2.5\%$ on each side). MCMC identifies a false peak in the lower energy region within which the response matrix is particularly ill-conditioned. This is reflected by the broad posterior confidence region (light blue region) around the posterior mean spectrum. This result is expected since Tik, PIDAL-O and MCMC all impose additional smoothness constraints on the spectrum.
%Moreover, GRAVEL noise variance parameters are set appropriately by incorporating realistic energy resolution to the detector response function \red{(Eq. \ref{eq:2.5}).} 

\begin{figure}[h!]
\centering
\includegraphics[width=1\columnwidth]{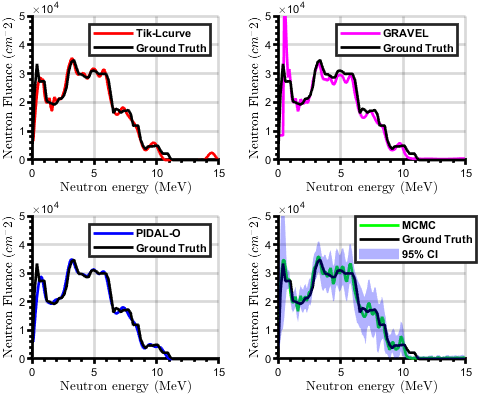}
\caption{Examples of unfolded spectra of the simulated $^{241}$AmBe neutron source ($5.10^7$ detection events per light output spectrum).}
\label{fig:4}
\end{figure}

\begin{figure}[h!]
\centering
\includegraphics[width=1\columnwidth]{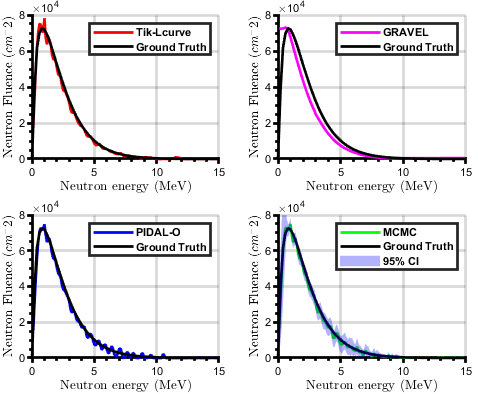}
\caption{Examples of unfolded spectra of the simulated $^{252}$Cf neutron source ($5.10^7$ detection events per light output spectrum).}
\label{fig:5}
\end{figure}

\begin{figure}[h!]
\centering
\includegraphics[width=1\columnwidth]{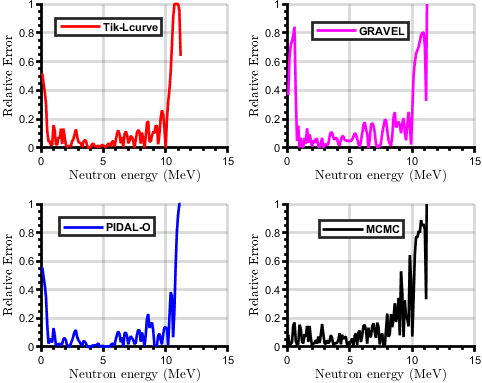}
\caption{Relative error plots of unfolded spectra of the simulated $^{241}$AmBe neutron source ($5.10^7$ detection events per light output spectrum) with respect to the Ground truth. Note PIDAL-O (PIDAL-Oracle) assumes full knowledge about ground truth spectra, so it serves as an estimate of the optimal unfolding algorithm and it is not attainable in actual experimental settings.}
\label{fig:6}
\end{figure}

\begin{figure}[h!]
\centering
\includegraphics[width=1\columnwidth]{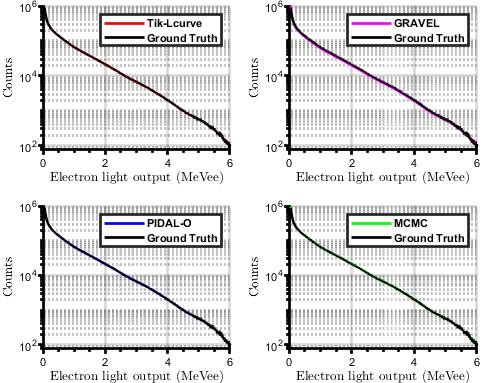}
\caption{Examples of light output spectra generated using the unfolded spectra of the simulated $^{241}$AmBe neutron source ($5.10^7$ detection events per light output spectrum) compared with ground truth light output. Note PIDAL-O (PIDAL-Oracle) assumes full knowledge about ground truth spectra, so it serves as an estimate of the optimal unfolding algorithm and it is not attainable in actual experimental settings}
\label{fig:7}
\end{figure}

Figs. \ref{fig:4} and \ref{fig:5} depict the unfolded spectra for the two continuous source ($^{252}$Cf and $^{241}$AmBe). Tik, GRAVEL, PIDAL-O and MCMC all show strong agreement with the ground truth spectrum. In addition, the credible intervals provided by the MCMC algorithm provides additional evidence about regions with higher uncertainty. Fig. \ref{fig:6} shows the relative error associated with the unfolded spectra with respect to ground truth for the $^{241}$AmBe  source. Fig. \ref{fig:7} shows the light output obtained as the convolution between the unfolded spectra and the response matrix compared to the ground truth light output. The four methods show very good agreement with the ground truth. This result illustrates one of the main challenges of the neutron unfolding problem, where several different unfolded spectra can lead to similar fits to the data to be deconvolved. Note that the relative error plots and generated light output plots for $^{252}$Cf lead to the same conclusions as those presented using $^{241}$AmBe, thus they are omitted here to reduce redundancy.

%@Haonan, it seems to compile now

We use the Spectral Angle Mapper (SAM) \cite{Keshava2002} between the unfolded spectrum ($\bf\hat{\bphi}$) and the known ground truth ($\bphi$) to quantify the unfolding performance of the different methods. Because the ground truth neutron spectra and response matrix have different neutron energy resolutions, we adopted SAM as opposed to standard Mean Square Error (MSE) as SAM is scale-invariant. Indeed, the SAM criterion relies on the spectral angle between $\bphi$ and $\hat{\bphi}$, which is small when $\bphi$ and $\hat{\bphi}$ present similar shapes. As a result, similar spectra lead to values of SAM close to $0$. The energy bounds listed in Table 1 were applied to the GRAVEL unfolded spectra to calculate the SAM.
\begin{eqnarray}
\label{eq:SAM}
SAM(\bphi,\hat{\bphi})=\arccos \left(\frac{\bphi\transp \bf\hat{\bphi}}{||\bphi||_2||\hat{\bphi}||_2}\right). 
\end{eqnarray}

Table \ref{table_3} summarizes all the SAMs which appear to be in agreement with the qualitative results as shown in Figs. \ref{fig:3} to \ref{fig:5}. Notably, MCMC, PIDAL and Tik all provided the competitive results based on SAM for the two continuous source, but MCMC automatically estimates the amount of regularization required from the data with additional credible interval.  

\begin{table}[h!]
%\begin{center}
\resizebox{1\columnwidth}{!}{
\begin{tabular}{|c|c|c|c|c|}
\hline
 \backslashbox{Neutron Source}{Method} & Tik &GRAVEL & PIDAL-O & MCMC \\
\hline
$\mathbf{DD}$ & 3.54  & 14.23  & 3.97 & 18.75  \\
\hline
$\mathbf{^{241}AmBe}$ & 6.26 & 4.6 13.30 & 6.29 & 5.13 \\
\hline
$\mathbf{^{252}Cf}$ & 2.97 & 4.73 14.14 &3.47 & 2.69 \\
\hline 
\end{tabular}}
%\end{center}
\caption{Spectral Angle Mapper (degrees) obtained using the different unfolding methods for the three sources ($5.10^7$ detection events per light output spectrum). Note PIDAL-O (PIDAL-Oracle) assumes full knowledge about ground truth spectra, so it serves as an estimate of the difficulty of the unfolding problem and it is not attainable in actual experimental settings}
\label{table_3}
\end{table}

%%--------------------------------------------------------------------%%
% The below Cos all except GRAVEL are cut off Results
%%-------------------------------------------------------------------%%
%\begin{table}[h!]
%\begin{center}
%\resizebox{1\columnwidth}{!}{
%\begin{tabular}{|c|c|c|c|c|}
%\hline
 %\backslashbox{Neutron Source}{Method} & Tik &GRAVEL & PIDAL-O & MCMC \\
%\hline
%$\mathbf{DD}$ & 30.77  & \red{32.69}  & 29.75 & 24.08  \\
%\hline
%$\mathbf{^{241}AmBe}$ & 3.96 & \red{4.6} & 3.20 & 5.25 \\
%\hline
%$\mathbf{^{252}Cf}$ & 2.89 & \red{4.73} &2.84 & 2.47 \\
%\hline 
%\end{tabular}}

In safeguards, security, and non-proliferation applications, it is often realistic to have a weak neutron signal that can be overwhelmed by an intense gamma-ray background \cite{Bourne2017}. Therefore, it is of considerable interest to examine the robustness of the algorithms as the number of detection event decreases (weak source and/or short integration time). We assess the robustness of the different algorithms using simulated data of $^{252}$Cf and  $^{241}$AmBe, for event counts ranging from $5\times 10^2$ up to $5\times 10^6$.
Note that for the most challenging scenarios, e.g., using only $5\times 10^2$ total counts across the $M=600$ light output bins, the average counts per bin fall below 1 for both $^{252}$Cf and  $^{241}$AmBe, with 480 empty bins on average for $^{241}$AmBe and 520 empty bins for $^{252}$Cf. This further motivates the use of the Poisson noise model in our unfolding procedure. The results are summarized in Fig. \ref{fig:6} and Table \ref{table_4}. Note that GRAVEL failed to converge for both sources at numbers of counts lower than $5\times 10^4$, which is denoted as N/A. 
 
As mentioned in Section \ref{subsec: proposed_approach}, PIDAL can be seen as a special case of the proposed hierarchical model where the hyperparameter $\delta$ is fixed as opposed to random. With appropriately tuned regularization parameters, Tik, PIDAL and MCMC demonstrated the competitive robustness against low counts. However, the proposed MCMC algorithm automatically adjusts this parameter and does not require exact knowledge about the ground truth.

%Due to the advantage of using sample mean as our final estimation, MCMC is the most robust algorithm in our experiments. 

% \begin{figure}[h!]
% \centering
% \includegraphics[width=1\columnwidth]{sam_final}
% \caption{Unfolding performance (Average SAM in degree) as a function of the total number of detection event for (a) $^{252}$Cf and (b) $^{241}$AmBe spectra. \red{In actual experiment, Tik (with L-curve Method), PIDAL-O and MCMC are all called 70 times to perform a log scale search to find the best tuning parameter(s) prior a full run}. PIDAL-O (PIDAL-Oracle) assumes full knowledge about ground truth spectra, so it serves as an estimate to the optimal unfolding algorithm and it is not attainable in actual experimental settings}
% \label{fig:8}
% \end{figure}

\begin{table}[h!]
\begin{center}
\resizebox{1\columnwidth}{!}{%
\begin{tabular}{|c|c|c|c|c|c|}
\hline
Neutron Source & Counts & Tik & GRAVEL & PIDAL-O & MCMC \\
\hline
	\multirow{5}{*}{$\mathbf{^{241}AmBe}$} & $5\times 10^6$ &8.99 (1.96) & 14.71 (2.99)
	%5.18 (0.67)%
	& 7.47 (0.77) & \textbf{7.99} (0.29) \\
 & $5\times 10^5$ &9.87 (0.46) & 15.81 (1.90) & 8.93 (0.86) & \textbf{9.89} (0.40)\\
  & $5\times 10^4$ &11.84 (0.49) & N/A & 10.96 (1.24) & \textbf{12.79} (0.65)\\
   & $5\times 10^3$ &15.25 (0.62)& N/A & 14.64 (1.54) & \textbf{17.06} (1.11)\\
    & $5\times 10^2$ &19.40 (3.75) & N/A & 17.18 (1.41) & \textbf{22.04} (2.61)\\
\hline
\hline
	\multirow{5}{*}{$\mathbf{^{252}Cf}$} & $5\times 10^6$ &4.69 (0.45) & 14.59 (1.02)
	% 9.54 (0.92) %
	& 4.54 (0.60) & \textbf{4.28} (1.12) \\
 & $5\times 10^5$ &5.05 (0.84) & 15.51 (1.60) & 5.78 (0.75) & \textbf{4.62} (1.06)\\
  & $5\times 10^4$ &7.06 (1.11) & N/A & 7.20 (1.02) & \textbf{6.33} (1.68)\\
   & $5\times 10^3$ &12.25 (1.14)& N/A & 10.35 (2.03) & \textbf{10.01} (2.26)\\
    & $5\times 10^2$ &16.97 (2.51) & N/A & 14.57 (3.22) & \textbf{22.73} (1.96)\\
\hline
\end{tabular}}
\caption{Unfolding performance (average SAM, in degree) as a function of the total number of detection event (best result per row in bold). Values in brackets represent standard deviations computed over 50 Monte Carlo realizations. Note PIDAL-O (PIDAL-Oracle) assumes full knowledge about ground truth spectra, so it serves as an estimate to the difficulty of the unfolding problem and it is not attainable in actual experimental settings.}
\end{center}

\label{table_4}
\end{table}

% \begin{table}[H]
% \begin{center}
% \resizebox{1\textwidth}{!}{%
% \begin{tabular}{|c|c|c|c|c|c|c|}
% \hline
% Neutron Source & Counts & MLE & GRAVEL & PIDAL & PIDAL-O & MCMC \\
% \hline
% 	\multirow{5}{*}{$\mathbf{^{241}AmBe}$} & 5000000 &39.07 (21.53) & 19.08 (0.78) & 7.08 (0.34) & 6.68 (0.51) & \textbf{7.52} (0.22) \\
%  & 500000 &53.64 (19.32) & 18.77 (0.80) & 8.65 (0.64) & 8.02 (0.75) & \textbf{9.83} (0.32)\\
%   & 50000 &67.67 (10.84) & N/A & 11.01 (0.76) & 10.51 (1.25) & \textbf{13.88} (0.56)\\
%    & 5000 &72.04 (6.29)& N/A &15.22 (1.23) & 13.84 (1.55) & \textbf{16.92} (0.63)\\
%     & 500 &77.17 (4.22) & N/A &18.95 (2.85) (5.65) & 18.55 (1.99) & \textbf{20.57} (2.56)\\
% \hline
% \hline
% \multirow{5}{*}{$\mathbf{^{252}Cf}$} & 5000000 &29.89 (6.15) &21.8 (0.55) &5.13 (0.20) &4.36 (0.69) & \textbf{5.44} (0.15) \\
%  & 500000 &45.56 (9.15) & 21.9 (1.21) &6.40 (0.43) &5.43 (0.72) & \textbf{7.15} (0.30)\\
%   & 50000 &57.13 (7.05)& N/A &8.53 (0.74) & 6.98 (1.05) & \textbf{9.83} (0.82)\\
%    & 5000 &63.01 (4.57)& N/A &11.90 (0.94) & 10.69 (2.12) & \textbf{14.88} (1.41) \\
%     & 500 &67.62 (3.01)& N/A &18.55 (3.01) & 14.66 (2.52) & \textbf{24.35} (2.34)\\
% \hline
% \end{tabular}}
% \end{center}

% \caption{Unfolding performance (average SAM, in degree) as a function of the total number of detection event (best result per row in bold). Values in brackets represent standard deviations computed over 50 Monte Carlo realizations.}
% \label{table_2}
% \end{table}

\noindent In practical applications, systematic errors in the unfolded spectra may arise because of an inaccurate calibration of the detector or a drift in the operating conditions, e.g.  temperature. In such cases, the presented methods are expected to exhibit a similar energy bias in the reconstructed spectrum since no strong prior information is incorporated into the algorithms. The unfolding of a known monoenergetic spectrum, e.g., from $^{137}$Cs, with suitable gamma-ray response matrix, could be used to mitigate and correct for such systematic errors. 
\noindent We implemented the Tik, PIDAL-O and the proposed MCMC unfolding algorithm in Matlab R2017b on an 2GHZ Intel processor with 6GB of RAM. The maximum number of iteration for Tik and PIDAL are fixed at 24000 but the algorithms generally converge and are stopped well before this number of iterations.  Within the MCMC algorithm, we generated sequentially 24000 samples (after the burn-in period of the sampler) for all the simulation results presented in this paper. Tik and PIDAL-O calls Tik and PIDAL to search for the best smoothing parameter. The tuning of hyperparameters of MCMC algorithm is done using WAIC (Watanabe-Akaike Information Criteria) \cite{gelman2014understanding}. We used the compiled version of GRAVEL available through RSICC (UMG package version 3.3). The average run time of the algorithms to analyze one spectrum is presented in Table \ref{tab:3}. As shown in Table \ref{tab:3}, the enhanced unfolding performance of the MCMC method comes with a significantly higher computational cost than Tik, GRAVEL and PIDAL (for a fixed value of the smoothing parameter) because the sequential nature of the sampler and the number of iterations required to estimate the posterior mean and credible intervals. Different choices of parameters for MCMC results in the significant discrepancy of run time for $^{241}$AmBe and $^{252}$Cf. In actual experiment, Tik (with L-curve Method) and  PIDAL-O  are called 70 times to perform a log scale search to find the best smoothing parameter prior a full run, while MCMC are called 6 times to perform a log scale search. However, it is worth noting that the hyperparameter selection procedure and the algorithm implemented has not been optimized for fast analysis, and it is possible to accelerate the method using C/C++ implementations.  

\begin{table}[h!]
\begin{center}
\resizebox{\columnwidth}{!}{%
\begin{tabular}{|c|c|c|c|c|}
\hline
 \backslashbox{Neutron Source}{Method} & Tik &GRAVEL & PIDAL & MCMC \\
\hline
$\mathbf{^{241}AmBe}$ & 0.38 & 900 & 0.45 & 83.39 \\
\hline
$\mathbf{^{252}Cf}$ & 0.71 & 60 & 0.53 & 40.42  \\
\hline 
\end{tabular}}
\end{center}

\caption{Average computational time to analyze one spectrum (in seconds) over 100 runs. Note all the reported time here excludes the additional parameter tuning time cost.}
\label{tab:3}
\end{table}

\section{Conclusions}
\label{sec:conclusion}
\noindent We have proposed a hierarchical Bayesian approach to solve the neutron spectrum unfolding problem, which differs from previous work \cite{choudalakis2012fully,kuusela2015statistical} by using an efficient constrained Hamiltonian Monte Carlo method and a hyper-prior on the hyper-parameter. The new MCMC algorithm shows improvement in performance compared to traditional approaches, such as Tik \cite{hansen1999curve}, GRAVEL \cite{matzke1994unfolding,MATZKE2002230,chen2014unfolding} and PIDAL \cite{figueiredo2010restoration} on simulated data ($^{252}$Cf and $^{241}$AmBe) in terms of accuracy with additional uncertainty evaluation through credible interval. This work further demonstrates the potential benefits of Bayesian methods for solving unfolding problems, because they provide a formalized manner in which to integrate existing prior knowledge within the estimation procedure. In this work, we have focused on synthetic data generated from reference neutron spectra and a known response matrix (ground truth available). In future work, the performance of the algorithm will be evaluated using measured data (simulated and measured response matrices) for organic scintillators. Efforts should in particular concentrate on robustness of the methods with respect to detector imperfections and background/spurious detections. Additional types of detectors with spectroscopic capability, e.g., Bonner sphere spectrometers, silicon telescopes, and superheated emulsions will also be investigated. The present unfolding method could also be coupled to classification algorithms to infer the type and amount of fissile material in unknown neutron sources, for nonproliferation and safeguarding applications. Approximate Bayesian methods will also be investigated for robust unfolding with reduced processing burden.   

\section*{Acknowledgments}
The authors acknowledge the support of the UK Royal Academy of Engineering under the Research Fellowship Scheme (RF201617/16/31) and the UK Engineering and Physical Sciences Research Council (EPSRC) via grant EP/S000631/1 and the MOD University Defence Research Collaboration (UDRC) in Signal Processing. This work was also funded in-part by the Consortium for Verification Technology under Department of Energy (DOE) National Nuclear Security Administration award number DE-NA0002534.

\bibliographystyle{IEEEtran}
\bibliography{mybibfile}

\end{document}